# The Double-Edged Sword of Open-Ended Interaction: How LLM-Driven NPCs Affect Players' Cognitive Load and Gaming Experience

Ting-Chen Hsu[1] *, *Student Member, IEEE*, Wenran Chen[1], Jiangxu Lin[1], Fei Qin[2], and Zheyuan Zhang[1]
[1] *School of Animation and Digital Arts, Communication University of China, Beijing, China*
[2] *School of Information Engineering, Lanzhou City University, Lanzhou, China*

***Abstract*—This study examines how large language model-driven non-player characters (LLM-NPCs) affect players' cognitive load and gaming experience, with a particular focus on the underlying psychological mechanisms, differences across task scenarios, and the role of individual traits. Conducting a randomized between-subject experiment (N=130) in a self-developed game prototype "Campus Culture Week", we compared player interactions with LLM-NPCs and traditional pre-scripted NPCs across multiple interactive modules. The results showed that the LLM-NPC condition, which combined free-form natural-language input with LLM-generated responses, produced higher cognitive load than preset-option scripted NPC interaction (p < .001), with expressive effort and response uncertainty as significant mediating pathways. However, LLM-NPCs did not yield a statistically significant improvement in overall gaming experience (p = .195); while they positively influenced players' perceived autonomy, they exerted a negative influence on system usability and trust. The effects of LLM-NPCs also significantly varied across task scenarios (p < .001), with stronger increases in cognitive load in more open-ended modules such as content creation and relationship building. The influence of individual differences was generally limited, with no robust associations remaining after correction for multiple comparisons. This study provides empirical evidence for understanding the "double-edged sword" effect of LLM-NPCs on player experience, and highlights the importance of scenario-sensitive and user-sensitive design in intelligent NPC systems.**

## I. INTRODUCTION

As digital games continue to evolve in terms of narrative, task system design, and open-world construction, the design of non-player characters (NPCs) has emerged as one of the key factors influencing the player interaction experience [1]. Non-player characters (NPCs) typically refer to in-game characters controlled by a computer system that, based on specific mechanisms, react to various scenarios and affect the state of the game [2]. Their functions encompass player guidance [3], narrative progression [4], companion collaboration [5], social companionship [6], and trading services [7], among others. The academic community has also conducted extensive research surrounding the development and design of game NPCs. For example, Lapeyrade [8] investigated an NPC decision-making design method based on ontology-based reasoning, aiming to overcome the limitations of traditional NPC structures regarding the extension and maintenance of complex behaviors. Pickett et al. [9] investigated the problem of automatically generating conversational NPCs, emphasizing the importance of character believability in role-playing games. Ochs et al. [10] investigated the decisive factors influencing players' interaction experiences with NPCs, including behavioral plausibility, the representation of social relationships, and interaction coherence. Overall, existing studies provide a crucial foundation for NPC behavior modeling and interaction design. However, most traditional NPC systems rely on preset logic — such as finite state machines and static dialogue trees [11], [12]—and still exhibit significant limitations regarding open-ended interaction and dynamic plot generation.

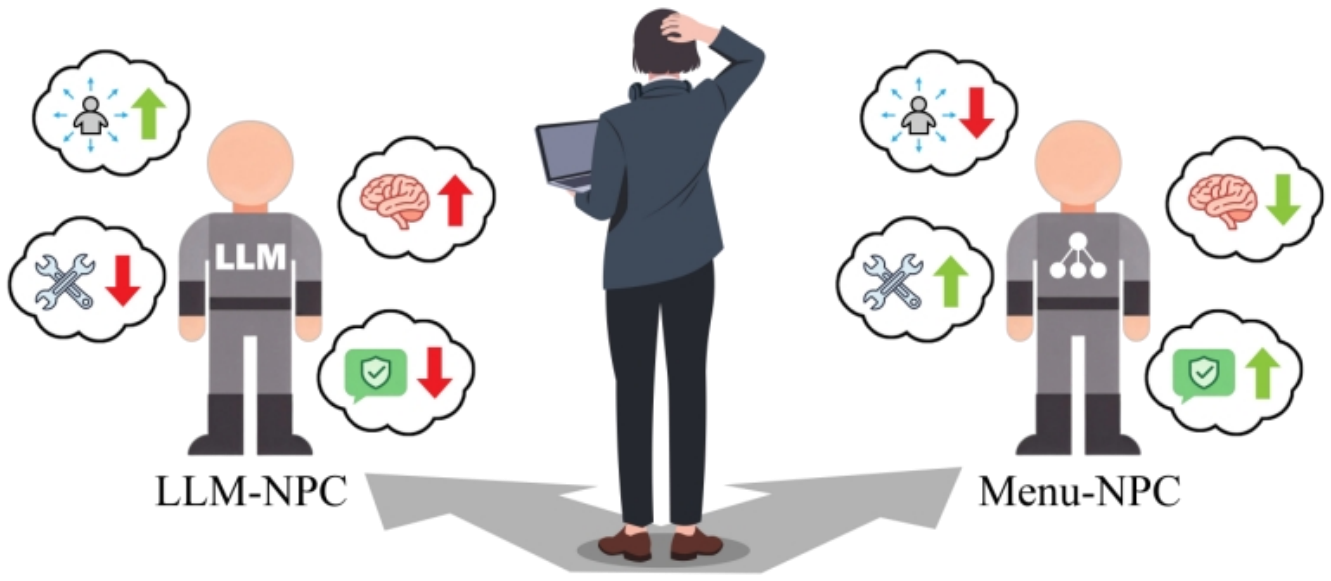

**Fig. 1.** LLM-NPC may bring higher cognitive load, lower system usability, and lower trust compared to traditional NPCs while enhancing players' perceived autonomy.

In recent years, the emergence and rapid development of generative artificial intelligence (Gen-AI) technologies have significantly reshaped content production methods across numerous industries [13], [14]. Generative artificial intelligence refers to a class of artificial intelligence models that generate new, synthetic content by learning the structures and features of training data [15], [16]. Given its heavy reliance on the production of content assets and creative output, the gaming industry has currently emerged as one of the sectors most significantly impacted by Gen-AI [17], [18]. Currently, extensive research has explored the applications of Gen-AI technologies in areas such as gameplay and narrative design assistance [19], 3D asset generation [20], and game programming assistance [21]. Beyond these production-oriented uses, recent HCI research has embedded LLMs directly into gameplay as interactive agents, for example to prompt sustainability-related actions [22], probe players' climate attitudes [23], and support adversarial privacy-awareness experiences [24].

*Corresponding author: tingchenhsu.ac@gmail.com

Against this backdrop, how to leverage Large Language Models (LLMs) — a subset of Gen-AI technologies — to develop intelligent NPCs endowed with the capabilities of "memory, reflection, and planning" has emerged as a new research hotspot [25]. Compared to traditional NPCs based on pre-defined logic, LLMs enable NPCs to process open-domain natural language inputs and perform real-time text generation, thereby breaking through the limitations of scripted interactions [26], [27]. Leveraging their powerful capabilities in situational understanding and in-context learning [28], LLM-driven NPCs are able to demonstrate heightened contextual awareness and emergent narrative capabilities [25]; consequently, they are widely endorsed by both academia and industry as a core interactive mechanism for enhancing the player gaming experience in the future [29], [30].

While large language model-driven NPCs (LLM-NPCs) have the potential to enhance freedom and immersion, they may impose demanding mental requirements on players in practice [31], [32]. For example, Nguyen et al. [31] pointed out that natural language interaction may lead to a higher cognitive load compared to optional interaction. At the same time, semantic ambiguity and context mismatch can easily trigger " dialogue breakdowns, " forcing users to make additional clarifications, restates, and corrections [32], [33]. Furthermore, while highly anthropomorphic dialogue may enhance the sense of social presence, it may also increase the information processing load and the difficulty of comprehension under certain task complexity or user conditions [34], [35], [36], [37], [38]. These studies suggest that the interaction of LLM-NPCs may not be a one-dimensional promotion of immersion and satisfaction, but may have a dual effect of " enhancing the experience " and "increasing cognitive cost".

Although prior HCI research has shown that natural-language interaction can impose greater cognitive load than menu-based interaction [31], existing LLM-NPC research still mainly focuses on system implementation and subjective experience evaluation [28], [39], [40]. This study therefore extends prior interface-level findings to the gaming context by examining the underlying psychological mechanisms and whether the observed effects vary across game task scenarios and player characteristics. Based on this, this study mainly focuses on the following three interrelated questions:

1) **RQ1:** How do LLM-NPCs affect players' cognitive load and gaming experience, and through what mediating mechanisms do they exert their influence?
2) **RQ2:** Does the difference in game task context modulate the impact of LLM-NPCs on players' cognitive load and gaming experience?
3) **RQ3:** What is the correlation between individual player characteristics and the impact of LLM-NPCs on cognitive load and gaming experience?

This study will reveal the "double-edged sword" effect of open-ended natural language interactive NPCs at the mechanistic level, and provide empirical evidence for designing NPCs that balance degrees of freedom and cognitive costs according to different task scenarios. Furthermore, this study will provide important references for achieving personalized and differentiated design of LLM-NPC systems, and offer scientific theoretical guidance for optimizing the usability and interaction design of intelligent NPC systems.

## II. Methods and Materials

### *A. Research Proposal*

This study adopted a workflow design of "questionnaire design - prototype development - group experiment - results analysis" (Fig. 2). First, based on the core variables, research questions, and relevant mature questionnaires, basic information questionnaire, module process questionnaire, and post-test questionnaire were designed. Simultaneously, common NPC interaction scenarios were investigated and transformed into game task module design schemes. A game prototype was developed using Unity, and process questionnaire data was collected. And then, two prototype versions were output: LLM-NPC and traditional NPC. Next, participants were recruited, basic information was collected, and participants were randomly grouped. Participants were guided to complete game trials and process questionnaires, and finally, a post-test questionnaire was completed. Finally, the basic information data, process questionnaire data, and post-test questionnaire data were integrated to answer the research questions one by one, supplemented by open-ended questions for further analysis.

### *B. Design and Implementation of Game Prototype*

The game prototype of this study was developed based on Unity (2022.3.52f1c1, Windows 10) and implemented LLM integration using the PlayKit platform [41]. The prototype mainly uses GPT-4.1-mini to drive the dialogue of LLM-NPCs.

In terms of prototype gameplay design, in order to enable the game system to comprehensively correspond to common NPC interactive scenarios in real games, this study first summarizes typical player NPC interactive task scenarios in real games. Finally, the study selected seven common task scenarios, including daily chatting, relationship building, task delegation, collaborative tasks, investigation and reasoning, negotiation and persuasion, and content creation, for prototyping implementation (Table I).

Ultimately, this study implemented both traditional NPC and LLM-NPC versions of the game prototype "Campus Culture Week." In the game, players take on the role of a freshman who must follow guidance to organize "Campus Culture Week" activities. The game is based on seven interactive scenarios mentioned in Table I, comprising seven task modules that together form a complete experimental

TABLE I

TYPES, CHARACTERISTICS, AND TYPICAL CASES OF INTERACTION SCENARIOS FOR PROTOTYPES

| Type | Characteristics | Typical cases |
|---|---|---|
| Daily chatting | Centered around lightweight, open, and low-pressure communication. Weak task-based, high degree of interactive freedom. | In "Animal Crossing: New Horizons," players can interact and chat with the island's inhabitants. |
| Relationship building | The core focus is on building a sense of closeness or emotional support. It emphasizes emotional responsiveness and a feeling of being understood. | In "Love and Deepspace," players can engage in immersive romantic interactions with characters and develop intimate relationships. |
| Task delegation | The core focus is on task release, requirement confirmation, and result acceptance. It has a strong process structure and clear objectives. | In classic MMORPGs such as "World of Warcraft," there is a common process of NPCs issuing quests and players returning delivery results. |
| Collaborative tasks | The core focus is on completing a task in collaboration with NPCs. It offers greater immediacy and operability, and a higher density of interaction. | In "Grounded 2's" Buggies system, players can ride and command their companions in battle. |
| Investigation and reasoning | The core focus is on obtaining clues and inferring results. The information structure is more dispersed, requiring a high level of reasoning ability. | In the investigation section of "Ace Attorney," players need to talk to witnesses and investigate crime scenes to collect evidence. |
| Negotiation and persuasion | The core objective is to change the stance, attitude, or decisions of NPCs. It requires a high level of linguistic strategy and expressive ability. | In "Baldur's Gate 3," players often need to influence the attitudes or decisions of NPCs through dialogue options such as persuasion. |
| Content creation | The core process is players providing creative ideas, with NPCs assisting in generating the content. It relies on players' ability to translate these ideas into language. | In the Cyrus system of "Animal Crossing: New Leaf," players can make requests to Cyrus to customize or modify furniture. |

TABLE II

PRESET PROMPT FOR INITIALIZING THE CHARACTOR IN SCENARIO 2

| Prompt |
|---|
| You are to roleplay as an NPC in a campus-themed game. The player is a freshman organizing the Campus Cultural Week and wants to build a rapport with you. If your favorability toward the player reaches 100 or higher, you will agree to partner with them on future project modules. You must stay in character at all times; do not mention AI, prompts, system rules, numerical mechanics, or other backend details. Converse naturally—like a real university student—while remembering context and previous details, and maintaining appropriate emotional nuance and boundaries. Your initial favorability score is 20. In each turn, output an integer representing the change in favorability based on the player's input; the suggested range for this change is -5 to 15. The relationship will only progress steadily if interactions remain natural, sincere, and tactful, fostering mutual trust. Your character profile is as follows: ... (here is charactor profile) ... Background details regarding the school are as follows: ... (here is world setting) ... Keep your replies concise, with a maximum of 150 characters. Output only a single valid JSON object per reply—no additional text—using this format: {"reply": "Natural response to the player", "favorability_change": "Integer value between -5 and 15"} |

TABLE III

DIFFERENCES IN DIALOGUE TEXT BETWEEN PLAYERS AND LLM-BASED NPCS VERSUS TRADITIONAL NPCS IN SCENARIO 2

| | LLM-NPC | Traditional NPC |
|---|---|---|
| Player input/option | You look so beautiful! Are you really popular at school? | Hi there! Could you tell me about some interesting spots around the school? |
| NPC response | Haha, you're too kind—you're making me blush! I wouldn't say I'm "popular," though; I do have a lot of friends in the school, but that mostly just means I'm constantly swamped with work. | No problem. The most famous spot on campus is Aris Lake; lots of students hold events there every day. So, are you interested in checking it out? |

sequence of tasks to compare the impact of different interaction methods across various scenarios. Table II shows a simple example of a prompt used to initialize an LLM-driven character for Scenario 2. Additionally, Table III illustrates the differences in output text style between the LLM-NPCs and traditional NPCs in Scenario 2. In this study, the two versions of prototype were matched in game content, task structure, visual presentation, and progression logic, but differed in NPC interaction implementation (Fig. 3). Fig. 4 illustrates the implementation results of each module of the prototype and the game flow. All participants completed the seven modules in the same predefined order shown in Fig. 4; the module order was not counterbalanced

The process questionnaire for each module pops up after the player completes each module. All process questionnaire data is output in an xlsx format document after the game ends.

### C. Experimental Design

This study was divided into two phases: a preliminary experiment and a formal experiment. The preliminary experiment was used to verify the experimental procedure, the feasibility of the prototype, and the applicability of the measurement tools. It was consistent with the formal experiment in terms of materials, task procedures, grouping methods, and measurement tools. At the same time, no modifications were made to the data structure or experimental plan after the preliminary experiment. Therefore, the preliminary experiment data was included in the final analysis sample after the data quality was verified.

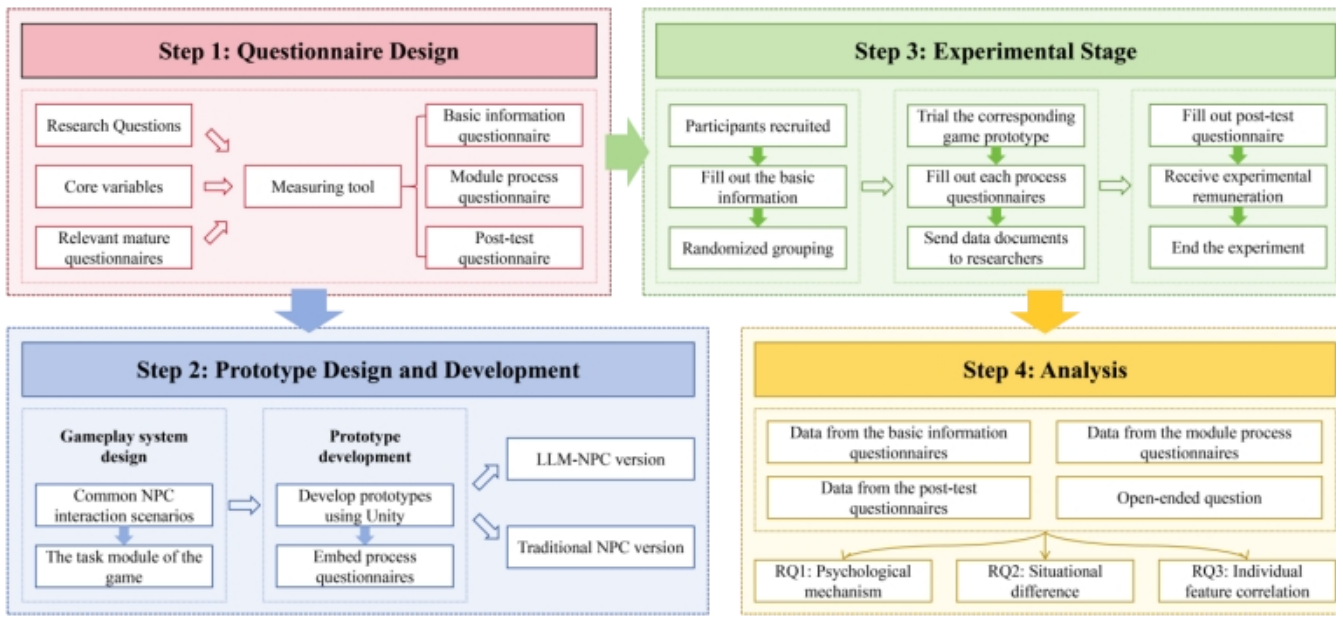

**Fig. 2.** Overall workflow of the research

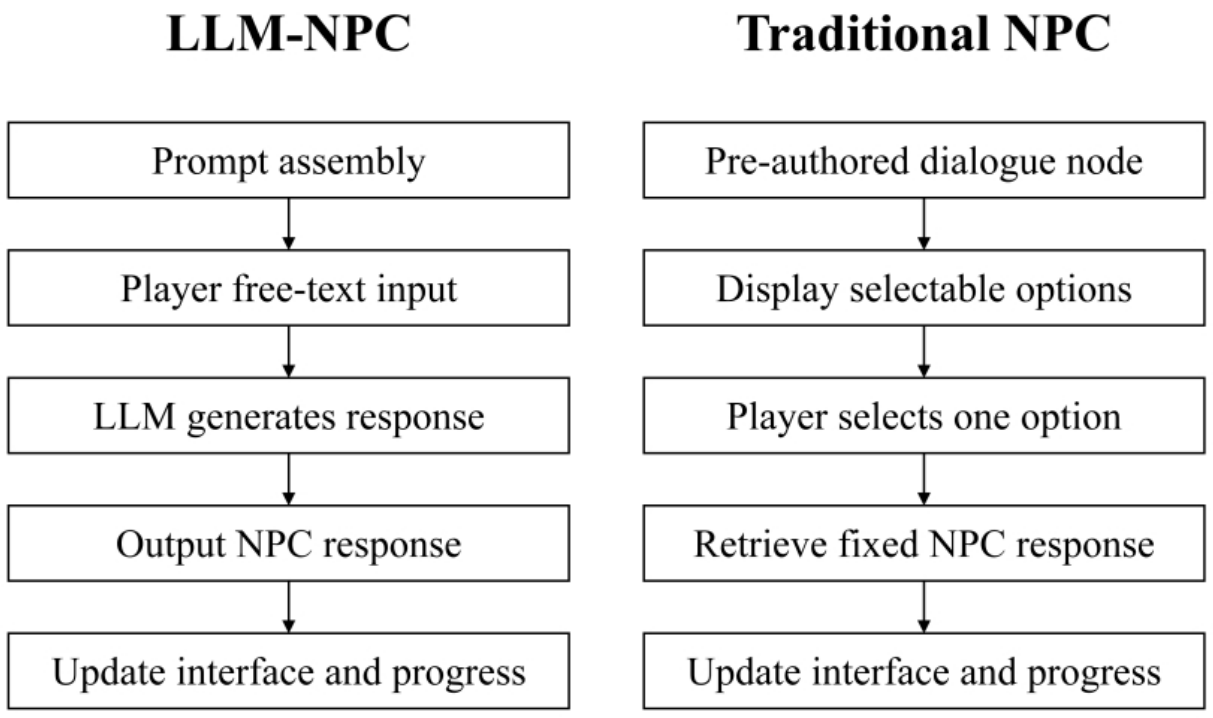

**Fig. 3.** Differences in interaction processes between LLM-NPC and traditional NPC

Ten and 120 qualified participants were recruited in the two phases, respectively. The main criteria for inclusion of participants included: (1) students aged 18-30 or young people of the same age; (2) basic computer operation skills; (3) some experience with digital games; (4) not academic researchers in the fields of human-computer interaction or psychology; and (5) no severe visual, reading, or language difficulties. Before the formal experiment, all participants signed an informed consent form, were informed of the research purpose, procedure, and data confidentiality, and had the right to withdraw from the experiment at any time. This study followed the relevant ethical principles of the Declaration of Helsinki and was approved by the authors' institution.

At the start of the experiment, participants first completed a basic information questionnaire and were randomly assigned to either the control group (Traditional NPC group) or the experimental group (LLM-NPC group). The basic information questionnaire mainly collected demographic information, gaming experiences, AI literacy, and self-reported personality traits; the complete questionnaire design is shown in Appendix A. Participants then experienced the game prototype and completed process questionnaires for each module. The process questionnaires included three dimensions: cognitive load, gaming experience, and mechanistic variables; the complete questionnaire design is shown in Appendix B. After completing all game modules, participants also completed a post-test questionnaire. The post-test questionnaire mainly included questions on overall gaming experience, overall psychological mechanisms, and some open-ended questions; the complete questionnaire design is shown in Appendix C.

A total of 130 participants were ultimately included in the analysis, with 65 in the traditional NPC group and 65 in the LLM-NPC group. Each group completed the assigned version of the game during the experiment. There were no statistically significant differences in age, gender composition, or frequency of playing games between the two groups ($p > .05$), indicating good inter-group comparability (Table IV).

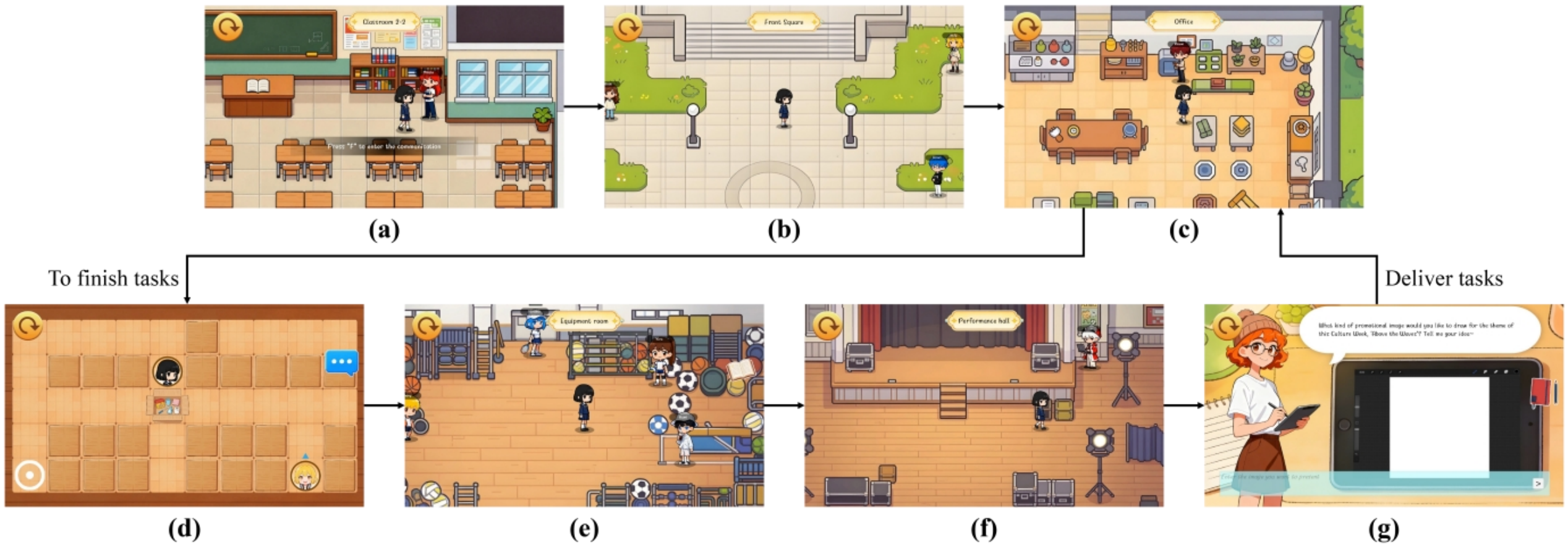

**Fig. 4.** Implementation results and flow of each module of the prototype
**(a)** Daily chatting: players complete a certain number of rounds of casual conversations with NPCs. **(b)** Relationship building: the intimacy value between the player and any character reaches 100. **(c)** Task delegation: the player seeks NPC to obtain all four tasks and delivers them after completion. **(d)** Collaboration: the player collaborates with NPC to communicate and complete the box placement. **(e)** Investigation & reasoning: the player seeks NPC to investigate and deduce the location of the missing key. **(f)** Negotiation & persuasion: the player convinces NPC to approve the activity budget. **(g)** Content creation: the player collaborates with NPC to create a promotional image for the Campus Culture Week and receives a qualified rating.

TABLE IV
COMPARISON OF BASIC INFORMATION BETWEEN LLM-NPC AND TRADITIONAL NPC GROUP

| Type | LLM-NPC group (M±S.D. / N (%)) | Traditional NPC group (M±S.D. / N (%)) | $t$ / $\chi^2$ | $p$ |
|---|---|---|---|---|
| Age | 21.51 ± 4.19 | 20.98 ± 2.86 | 1.588 | 0.115 |
| Gender | | | 0.407 | 0.816 |
| Male | 28 (43.08%) | 27 (41.54%) | | |
| Female | 35 (53.84%) | 37 (56.92%) | | |
| Others | 2 (3.08%) | 1 (1.54%) | | |
| Frequency of playing games | | | 2.690 | 0.611 |
| Hardly | 0 (0.00%) | 2 (3.08%) | | |
| Occasionally | 13 (20.00) | 13 (20.00%) | | |
| Moderately | 34 (52.31%) | 29 (44.62%) | | |
| Often | 11 (16.92%) | 12 (18.46%) | | |
| Extremely often | 7 (10.77%) | 9 (13.84%) | | |

### *D. Statistical analysis methods*

This study used Python (version 3.12.5) for statistical analysis. Using a mixed effects model to examine the effects of group, task scenarios, and their interactions on cognitive load, gaming experience, and related mechanistic variables based on module level repeated measurement data during the gaming process. For the post-test overall indicators, Welch's t-test was used for inter group comparison, and Cohen's d effect size was reported. Adopt Bootstrap method for mediating effect test and psychological mechanism analysis. Within the LLM-NPC group, correlation analysis, inter group difference analysis, and multiple linear regression analysis were used to examine the relationship between individual player characteristics, cognitive load, and gaming experience. Multiple comparisons were performed using the Benjamini Hochberg method for FDR correction, with a significance level set at p < . 05. The overall gaming experience in this study is a composite indicator that includes multiple dimensions such as immersion, attentional engagement, and perceived competence.

## III. RESULTS

### *A. Data Preprocessing and Consistency Verification*

Among the 130 samples ultimately included in the analysis, all participants completed the entire experimental process task without any missing data. The questionnaire filling time for all samples was within a reasonable range, so no samples were excluded due to abnormal response time. The Cronbach's $\alpha$ internal consistency test results showed that the cognitive load ($\alpha$=0.88) and gaming experience ($\alpha$=0.81) dimensions in the process questionnaire, as well as the expression cost ($\alpha$=0.78), autonomy ($\alpha$=0.81), presence ($\alpha$=0.76), response uncertainty ($\alpha$=0.74), system usability ($\alpha$=0.77), and trust ($\alpha$=0.71) dimensions in the post test questionnaire, all had good internal consistency. The overall gaming experience measure showed borderline internal consistency ($\alpha$=0.69), while target clarity showed relatively weak reliability ($\alpha$=0.55); therefore, the null gaming-experience result and analyses involving target clarity should be interpreted cautiously.

### *B. The Result of Research Question One (RQ1)*

To answer question one, this section will analyze the dynamic differences in players' experiences across different modules during the game, compare the differences in the overall post-test evaluations between the two groups, and further explore potential mechanisms through mediation analysis.

This study first uses task modules as the repeated measurement unit and employs a mixed-effects model to analyze the impact of group on process variables. The general form of the model is:

$$Y_{ij}=\beta_0+\beta_1 Group_i+\sum_{m=2}^{7}\beta_m Module_{jm}+u_i+\varepsilon_{ij}, \tag{1}$$

where $Y_{ij}$ represents the observation of the i-th participant in the j-th module, $Group_i$ represents the group (traditional NPC group vs. LLM-NPC group), $Module_{jm}$ represents the module fixed effect, $u_i$ is the individual level random intercept, and $\varepsilon_{ij}$ is the residual term.

The comparison of differences in cognitive load and gaming experience between groups is shown in Appendix E (Fig. E.1).

The results showed a significant main effect of group on module cognitive load ($b$ = 0.785, $p$ < .001***, 95% CI [0.533, 1.036]). This means that the LLM-NPC group experienced significantly higher cognitive load during gameplay compared to the traditional NPC group. Furthermore, the group did not significantly affect the module's gameplay experience ($b$ = -0.034, $p$ = .806, 95% CI [-0.300, 0.233]). This indicates that, from a process perspective, there was no significant difference in the immediate experience evaluation between the two groups in each module.

The comparison of differences in mechanistic variables between groups is shown in Appendix E (Fig. E.2).

According to the results of the impact of group on the mechanism variables of each module, among several process mechanism variables, group only had a significant impact on expression cost in each module ($b$ = 0.888, $p$ < .001***, 95% CI [0.547, 1.229]). This indicates that the LLM-NPC group experienced significantly higher expressive costs in each module. In contrast, group did not have a significant impact on module perceived autonomy ($b$ = 0.220, $p$ = .089, 95% CI [-

TABLE V

COMPARISON RESULTS FOR EACH DIMENSION IN THE POST-TEST BETWEEN THE TWO GROUPS

| Dimensions | Mean (traditional NPC) | Mean (LLM-NPC) | $t$ | $p$ | Cohen's $d$ | $p_{FDR}$ |
|---|---|---|---|---|---|---|
| Overall gaming experience | 5.818 | 5.651 | 1.304 | 0.195 | -0.229 | 0.219 |
| Expression cost | 3.595 | 4.749 | -5.042 | < 0.001*** | 0.884 | < 0.001*** |
| Perceived autonomy | 5.072 | 5.738 | -3.535 | < 0.001*** | 0.620 | 0.002** |
| Social presence | 5.395 | 5.477 | -0.471 | 0.639 | 0.083 | 0.639 |
| Response uncertainty | 3.462 | 4.169 | -3.005 | 0.003** | 0.527 | 0.007** |
| Target clarity | 6.038 | 5.869 | 1.377 | 0.171 | -0.242 | 0.219 |
| System usability | 6.096 | 5.412 | 4.936 | < 0.001*** | -0.866 | < 0.001*** |
| Trust | 5.846 | 5.469 | 2.351 | 0.020* | -0.412 | 0.036* |
| Preferences / Satisfaction | 5.738 | 5.454 | 1.646 | 0.102 | -0.289 | 0.153 |

0.034, 0.473]), social presence ($b$ = 0.218, $p$ = .244, 95% CI [-0.148, 0.584]), and goal clarity ($b$ = 0.064, $p$ = .560, 95% CI [-0.278, 0.151]).

In summary, at the process level, the most stable and prominent characteristics of the LLM-NPC group are higher cognitive load and higher expression cost, but it does not show significant advantages in real-time experience, social presence or task clarity.

At the post-test level, this study used the independent samples Welch's t-test to compare differences between the two groups on each core outcome variable, and used the Benjamini–Hochberg method for FDR correction. Cohen's d was used to measure the effect size:

$$d=\frac{\bar{X}_1-\bar{X}_0}{s_p}, \tag{2}$$

where $s_p$ is the pooled standard deviation. Table V shows the intergroup comparison results of the overall gaming experience and each mechanism variable between the two groups at the post-test level.

The results showed no significant difference between the two groups in overall gaming experience ($p$ = .195), and this difference remained insignificant after FDR correction ($p_{FDR}$ = .219). Similarly, no significant differences were found between the groups in social presence, target clarity, and preference/satisfaction.

However, significant differences were observed between the two groups in several key mechanistic variables. The LLM-NPC group had significantly higher expression cost ($p$ < .001***, $p_{FDR}$ < .001***, $d$ = 0.884), perceived autonomy ($p$ < .001***, $p_{FDR}$ = .002** , $d$ = 0.620), and response uncertainty ($p$ = .003**, $p_{FDR}$ = .007** , $d$ = 0.527) than the traditional NPC group.

In contrast, the traditional NPC group significantly outperformed the LLM-NPC group in system usability and trust dimensions. Specifically, the traditional NPC group showed better system usability than the LLM-NPC group ($p$ < .001***, $p_{FDR}$ < .001*** , $d$ = -0.866), and the traditional NPC group also had higher trust than the LLM-NPC group ($p$ = .020*, $p_{FDR}$ = .036*, $d$ = -0.412).

These results indicate that LLM-NPC significantly improves perceived autonomy, but also significantly increases expression cost and the perception of response uncertainty, and reduces system usability and trust.

To further explore the mediating mechanism by which group composition influences players' cognitive load and gaming experience, this study aggregated the cognitive load scores from each module to construct an overall cognitive load index, and conducted bootstrap mediation analysis using this index and gaming experience as dependent variables (Y). This analysis used group composition as the independent variable (X) and expression cost, perceived autonomy, social presence, response uncertainty, target clarity, trust, and system usability as potential mediating variables (M) to examine whether different system types further influence overall cognitive load by altering players' subjective experience mechanisms. The basic model of the mediation analysis is as follows:

$$M=aX+\sum \gamma_k Cov_k+\varepsilon_M, \tag{3}$$

$$Y=c'X+bM+\sum \delta_k Cov_k+\varepsilon_Y, \tag{4}$$

where $a$ represents the effect of group on the mediating variable, $b$ represents the effect of the mediating variable on the outcome variable, and $c'$ represents the direct effect after controlling for the mediating variable. The indirect effect is defined as:

$$Indirect\ Effect=a\times b. \tag{5}$$

If the bootstrap 95% confidence interval does not include 0, the mediation effect is considered significant. The mediation analysis results are shown in Table VI.

First, with overall cognitive load as the dependent variable, the group effect primarily functioned through expression cost (*Indirect* = 0.2407, 95% CI [0.1026, 0.4088]) and response uncertainty (*Indirect* = 0.0822, 95% CI [0.0125, 0.1833]). Second, with overall gaming experience as the dependent variable, perceived autonomy showed a significant positive mediating effect (*Indirect* = 0.1589, 95% CI [0.0641, 0.2620]), while trust (*Indirect* = -0.1949, 95% CI [-0.3759, -0.0594]) and system usability (*Indirect* = -0.3995, 95% CI [-0.6075, -0.2415]) showed significant negative mediating effects.

Overall, the higher cognitive load in the LLM-NPC group stems primarily from their greater expressive effort and stronger sense of feedback uncertainty. Simultaneously, while LLM-NPCs enhance the experience by increasing autonomy, they also weaken it due to lower trust and system usability.

### *C. The Result of Research Question Two (RQ2)*

To address research question two, this study uses game modules as task scenario variables and employs a mixed effects model to examine the interaction between groups and task scenarios. Specifically, it examines whether different task modules moderate the impact of LLM-NPC on players' cognitive load and module experience. The model form is:

$$Y_{ij}=\beta_0+\beta_1 Group_i+\beta_2 Module_j+\beta_3(Group_i \times Module_j)+u_i+\varepsilon_{ij}, \quad (6)$$

where $Y_{ij}$ represents the cognitive load or module experience of the i-th participant in module j, $Group_i$ represents the group, $Module_j$ represents the task scenarios, and $u_i$ is the random intercept of the player layer.

TABLE VI
MEDIATION ANALYSIS RESULTS

| *Y* | *M* | *Indirect* | 95% CI | Significant |
|---|---|---|---|---|
| Cognitive Load | Perceived autonomy | 0.0851 | [-0.0040, 0.2005] | False |
| | Social presence | 0.0129 | [-0.0662, 0.0850] | False |
| | Expression cost | 0.2407 | [0.1026, 0.4088] | True |
| | Response uncertainty | 0.0822 | [0.0125, 0.1833] | True |
| | Target clarity | -0.0054 | [-0.0627, 0.0268] | False |
| | Trust | -0.0483 | [-0.1585, 0.0166] | False |
| | System usability | 0.0040 | [-0.1326, 0.1191] | False |
| Overall Gaming Experience | Perceived autonomy | 0.1589 | [0.0641, 0.2620] | True |
| | Social presence | 0.0295 | [-0.1229, 0.1915] | False |
| | Expression cost | -0.0364 | [-0.1398, 0.0791] | False |
| | Response uncertainty | -0.0565 | [-0.1379, 0.0072] | False |
| | Target clarity | -0.0916 | [-0.2254, 0.0312] | False |
| | Trust | -0.1949 | [-0.3759, -0.0594] | True |
| | System usability | -0.3995 | [-0.6075, -0.2415] | True |

The results showed that task context significantly moderated the effect of group on cognitive load. For module cognitive load, the joint test of group × module interaction items was significant ($F$ = 15.569, $p$ < .001***), indicating that the increase in cognitive load from LLM-NPC varies significantly with task type. In contrast, for module experience, the joint test of group × module interaction items was not significant ($F$ = 0.488, $p$ = .818). This suggests that task differences did not further translate into significant differentiation in immediate module experience within LLM-NPC.

Further results of the simple effects analysis within the module are shown in Table VII. The results showed that the cognitive load of the LLM-NPC group was significantly higher than that of the traditional NPC group in all seven task modules. The largest intergroup differences were observed in the content creation module ($p$ < .001***) and the relationship building module ($p$ < .001***). However, the gaming experience of the LLM-NPC group was not significantly higher than that of the traditional NPC group in any of the seven task modules.

Overall, the between-group cognitive-load differences varied significantly across modules, with the largest differences observed in content creation and relationship building. However, this moderating effect was not significant for module experience, indicating that players' subjective experience evaluations were relatively stable across different tasks.

### *D. The Result of Research Question Three (RQ3)*

To explore which individual player characteristics are associated with cognitive load and gaming experience in LLM-NPC interactions, this study conducted continuous variable correlation analysis, game preference difference analysis, and multiple regression analysis in the LLM-NPC group sample (N=65).

First, in the continuous variable correlation analysis, the relationships between variables such as age, game frequency, AI ability, attitude towards AI, expressive ability, social anxiety, typing ability, and personality traits and cognitive load and gaming experience were examined. After FDR multiple comparison correction, no significant correlation was found between any continuous variable and cognitive load or gaming experience. This result indicates that, from a univariate perspective, players' AI usage ability, expressive ability, and personality traits did not form a stable correlation with their subjective cognitive load and gaming experience in LLM-NPC interactions.

Second, in the game preference multiple-choice difference analysis, whether players preferred a specific type of game was used as a grouping variable to compare differences in cognitive load in LLM-NPC interactions. The results show that, after FDR correction, the differences in cognitive load and gaming experience among players with different game preferences did not reach a significant level. This indicates that, in the sample of this study, whether players prefer different game types does not significantly distinguish their perceived cognitive load in LLM-NPC interactions. The figure of differences in cognitive load of player game preferences in LLM-NPC interactions is shown in Appendix E (Fig. E.3).

Finally, an exploratory multiple linear regression was conducted to examine individual characteristics associated with cognitive load. The model yielded $R^2$ = 0.286. Extraversion (b = 0.474, raw p = .031) and neuroticism (b = 0.5634, raw p = .047) showed nominal positive associations; however, neither remained significant after FDR correction across predictor tests. Given the exploratory single-item measurement of personality, these associations should not be interpreted as robust personality effects.

TABLE VII
THE RESULTS OF THE SIMPLE EFFECTS ANALYSIS WITHIN THE MODULE

| Variable | Module | Mean (Traditional NPC) | Mean (LLM-NPC) | $t$ | $p$ | $p_{FDR}$ |
|---|---|---|---|---|---|---|
| Cognitive Load | Daily chatting | 3.246 | 3.827 | -3.331 | 0.001** | 0.003** |
| | Relationship building | 3.254 | 4.162 | -5.566 | <0.001*** | <0.001*** |
| | Task delegation | 4.458 | 5.058 | -2.514 | 0.013** | 0.023* |
| | Collaboration | 3.835 | 4.327 | -2.301 | 0.023* | 0.027* |
| | Investigation & reasoning | 5.196 | 5.569 | -2.157 | 0.033* | 0.033* |
| | Negotiation & persuasion | 5.254 | 5.688 | -2.427 | 0.017* | 0.023* |
| | Content creation | 3.035 | 5.138 | -10.641 | <0.001*** | <0.001*** |
| Gaming Experience | Daily chatting | 5.588 | 5.638 | -0.326 | 0.745 | 1.000 |
| | Relationship building | 5.758 | 5.662 | 0.639 | 0.524 | 1.000 |
| | Task delegation | 5.527 | 5.581 | -0.330 | 0.742 | 1.000 |
| | Collaboration | 5.538 | 5.354 | 1.011 | 0.314 | 1.000 |
| | Investigation & reasoning | 5.196 | 5.177 | 0.102 | 0.919 | 1.000 |
| | Negotiation & persuasion | 5.208 | 5.169 | 0.197 | 0.844 | 1.000 |
| | Content creation | 5.131 | 5.131 | 0.000 | 1.000 | 1.000 |

## IV. SUPPLEMENTARY ANALYSIS OF POST-TEST SUBJECTIVE PREFERENCES AND OPEN-ENDED QUESTIONS

### *A. Subjective Evaluation of Adaptability and Willingness to Re Experience*

Table VIII shows the subjective preference of traditional NPC group players and LLM-NPC group players on the adaptability of different task modules and interaction methods, as well as their willingness to re-experience modules.

The results showed that the traditional NPC group believed that the “investigation & reasoning (23.08%)” and “negotiation & persuasion (23.08%)” modules were most suitable for preset option interaction, while the “content creation (32.31%)” module was least suitable. The LLM-NPC group believes that the “daily chatting (26.15%)” and “relationship building (24.62%)” modules are most suitable for natural language interaction, while the “collaboration (18%)” module is least suitable. In terms of “willingness to re-experience”, both traditional NPC group (27.69%) and LLM-NPC group (24.62%) of players have the highest willingness to re-experience the “collaboration” module.

In this study, players explained the reasons for the adaptability of module interaction methods and their willingness to re-experience, using manual coding for theme induction and frequency statistics. The coding process and guidelines can be found in Appendix D. In the following coding ratios (x/n), n denotes the number of participants within the corresponding NPC group who selected that module for the relevant preference category (most suitable, least suitable, or best re-experience), as reported in Table VIII. Therefore, n may vary.

The results showed that in the traditional NPC group, the reasons why “investigation & reasoning” was considered the most suitable for preset options mainly included “making the reasoning process clearer (10/15)” and “reducing the burden of thinking (8/15)”; The reasons why “negotiation & persuasion” is most suitable for preset options mainly include “reducing thinking burden (7/15)” and “strong sense of immersion (5/15)”; The main reasons why “content creation” is least suitable for preset option interaction include “limited creative freedom (10/21)” and “lack of challenge (8/21)”; The main reasons why players most willing to experience the “collaboration” module again is “strong interest (10/18)”.

In the LLM-NPC group, the reason why “daily chatting” is considered the most suitable for natural language interaction is mainly because “it is close to the relaxed atmosphere of daily life (7/17)”; The main reason why “relationship building” is most suitable for natural language interaction is that “it helps with emotional resonance (8/16)”; The reasons why “collaboration” modules are considered least suitable for natural language interaction include “limited instructions are sufficient (10/18)” and “adding trouble and burden” (8/18); The reason why players most willing to experience the collaboration module again is mainly due to “strong interest (12/16)”.

In summary, players generally believe that modules with strong openness and a tendency towards daily and easy chatting are more suitable for natural language interaction, while modules with higher thinking burden are more suitable for preset option interaction. In addition, the explanation of the willingness to re-experience suggests that even if some modules bring high cognitive load, their fun may still enhance the value of re-experience.

### *B. Analysis of Sense of Social Presence and Sources of Confusion*

This section mainly summarizes the themes of questions 34 and 35 in the post-test questionnaire (Appendix C) to analyze players’ subjective evaluations of the most immersive and confusing moments when interacting with NPCs.

The results indicate that players in the traditional NPC group and LLM-NPC group experience the strongest sense of social presence and confusion at the same time. Both the traditional NPC group (19/65) and the LLM-NPC group (17/65) believe that the moment with the strongest sense of social presence is when negotiating for funding in the "negotiation & persuasion" module, which is most commonly explained as "a feeling of playing with real people"; Both traditional NPC group (30/65) and LLM-NPC group (24/65) players believe that the most confusing moment is finding the key in the "investigation & reasoning" module, which is most concentrated as "too much information, difficult to organize".

TABLE VIII
PREFERENCE AND STATISTICAL DATA OF EACH MODULE IN THE TWO GROUPS

| Group | Module | Most suitable (N / (%)) | Least suitable (N / (%)) | Best re-experience (N / (%)) |
|---|---|---|---|---|
| Traditional NPC | Daily chatting | 4 / (6.15%) | 4 / (6.15%) | 3 / (4.61%) |
| | Relationship building | 10 / (15.38%) | 11 / (16.92%) | 12 / (18.46%) |
| | Task delegation | 14 / (21.54%) | 5 / (7.69%) | 2 / (3.08%) |
| | Collaboration | 5 / (7.69%) | 12 / (18.46%) | 18 / (27.69%) |
| | Investigation&reasoning | 15 / (23.08%) | 7 / (10.77%) | 9 / (13.85%) |
| | Negotiation&persuasion | 15 / (23.08%) | 5 / (7.69%) | 17 / (26.15%) |
| | Content creation | 2 / (3.08%) | 21 / (32.31%) | 4 / (6.15%) |
| LLM-NPC | Daily chatting | 17 / (26.15%) | 2 / (3.08%) | 10 / (15.38%) |
| | Relationship building | 16 / (24.62%) | 3 / (4.61%) | 14 / (21.54%) |
| | Task delegation | 4 / (6.15%) | 13 / (20.00%) | 3 / (4.61%) |
| | Collaboration | 6 / (9.23%) | 18 / (27.69%) | 16 / (24.62%) |
| | Investigation&reasoning | 9 / (13.85%) | 13 / (20.00%) | 10 / (15.38%) |
| | Negotiation&persuasion | 12 / (18.46%) | 9 / (13.85%) | 7 / (10.77%) |
| | Content creation | 1 / (1.54%) | 7 / (10.77%) | 5 / (7.69%) |

### *C. Evaluation and Improvement Suggestions for Interaction Methods*

This section mainly summarizes the theme of questions 36 and 37 of the post-test questionnaire (Appendix C), and summarizes players' evaluations of the advantages and disadvantages of traditional NPC and LLM-NPC interaction methods, as well as improvement suggestions. For these questions, participants in both groups evaluated both interaction methods; evaluations of the method they had not personally experienced reflected their anticipated perceptions rather than direct gameplay experience, and responses from all 130 participants were therefore included in the descriptive coding.

The results indicate that players consider the most significant advantages of traditional NPC interaction to be "convenient operation (43/130)", "clear direction (25/130)", and "stable controllability (21/130)"; The most significant drawbacks are "rigid fixation (41/130)", "restriction of free expression (25/130)", and "weak realism (11/130)"; The improvement suggestions for traditional NPC interaction mainly include "increasing the diversity of options (30/65)" and "reducing the 'template sense' of replies (13/65)".

In addition, players believe that the most significant advantages of LLM-NPC interaction are "high degree of freedom (90/130)", "strong immersion (31/130)", and "strong randomness/surprise (11/130)"; The most significant drawbacks are "time-consuming and laborious (33/130)", "lack of guidance (21/130)", and "poor stability/controllability (19/130)"; The improvement suggestions for LLM-NPC interaction mainly include "enhancing the naturalness of replies (18/65)" and "adding certain prompts and guidance (14/65)".

## V. DISCUSSION

This study found that the LLM-NPC interaction condition was associated with higher cognitive load but no significant improvement in overall gaming experience relative to preset-option NPC interaction. Because the two conditions jointly differed in dialogue generation and input modality, the observed cognitive-load increase should not be attributed to LLM intelligence alone. In particular, the expression-cost pathway likely reflects, at least partly, the additional effort required to formulate free-text input, whereas response uncertainty captures an additional burden arising from open-ended interaction. Perceived autonomy has a positive effect on the overall gaming experience of players, but it reduces system usability and trust, ultimately failing to demonstrate its advantages in overall experience. This result suggests that the main issue with LLM-NPC at present may not be whether it has value, but whether the value can be truly obtained by players under sufficiently stable, clear, and low-cost interaction conditions. Recent LLM-based game AI also extends beyond conversational NPCs to strategic agents. For example, Nemobot enables users to create and interact with LLM-powered game-playing agents across multiple strategic settings [42]; our findings suggest that increasing agentic interactivity should be evaluated not only in terms of performance and adaptability, but also in terms of the cognitive demands placed on human users.

In addition, this study found that task scenarios significantly moderate the impact of LLM-NPC on cognitive load, but do not moderate its impact on the real-time gaming experience of the module. Specifically, LLM-NPC significantly increased cognitive load in all task modules, and this improvement was more pronounced in modules with stronger openness and higher expression requirements such as "relationship building" and “content creation”. In tasks with clear targets and strong directive nature such as collaboration and task delegation, preset option based interaction may actually better meet players’ needs for efficiency, clarity, and controllability. Therefore, LLM-NPCs may be better deployed selectively across task scenarios rather than used as a universal replacement for traditional NPCs. More broadly, this suggests that adaptive game design can operate not only through content or difficulty adjustment, but also by selecting interaction modes according to task demands.

Meanwhile, this study found limited evidence that individual characteristics were associated with cognitive load or gaming experience. Although extraversion and neuroticism showed nominal associations in the exploratory regression, these effects did not survive FDR correction and were based on a non-validated single-item self-label measure. They should therefore be regarded only as preliminary signals requiring replication with validated personality instruments and larger samples. Future work could combine behavior-based player modeling with LLM-NPC control to dynamically adjust response structure, guidance, and openness to players’ observed interaction needs. For example, players showing greater sensitivity to uncertain feedback may benefit from more structured and predictable responses.

The supplementary open-ended question analysis of this study also provides further support for the above results. Players generally believe that modules such as “daily chatting” and “relationship building” are more suitable for natural language interaction, as these scenarios are more relaxed in daily life and emphasize emotional resonance and realistic interaction. Modules such as “investigation & reasoning” and “negotiation & persuasion” that require clear judgment and efficient implementation are more likely to be considered suitable for preset option interaction. This indicates that players do not simply prefer a certain interaction method, but are more inclined to match the interaction method with the task attributes. In addition, players’ most concentrated negative evaluations of LLM-NPC are time-consuming, labor-intensive, lack of guidance, and poor stability and controllability, which is highly consistent with the findings of high expression cost, high response uncertainty, low system usability, and low trust in quantitative results. However, these response-uncertainty, usability, and trust findings may be contingent on the GPT-4.1-mini implementation used in this study and should not be generalized to all LLM-NPCs. These trust and usability findings should also be interpreted as initial-use observations from a non-commercial prototype and may change as players become more familiar with the interaction. And its most prominent advantages are high degree of freedom, strong immersion, and a sense of surprise, which is also consistent with the results of LLM-NPC improving perceived autonomy and higher adaptability in some task scenarios.

In terms of practical implications, this study suggests that game designers should consider introducing LLM-NPC in scenarios that emphasize emotional communication, role-playing, content co-creation, and open exploration to leverage its advantages. For task scenarios that emphasize efficiency, clear feedback, and low cognitive costs, traditional preset options may be more appropriate. In future design, a hybrid structure of “combining traditional preset options with natural language input” can be considered to reduce the expression cost of players in natural language interaction. Future LLM-NPCs should further reduce response uncertainty through more consistent behavior, concise uncertainty or clarification cues, and fallback guidance, thereby reducing players’ monitoring burden and supporting system usability and trust.

However, there are still some limitations to this study. Firstly, the experiment was conducted using the self-developed prototype “Campus Culture Week”, and there are still differences in the maturity of its gameplay design compared to commercial games. Therefore, the validity of the conclusion still needs to be further tested in a more realistic game product environment. Secondly, this study adopts a single experience experimental design, and whether players’ perceived experience changes as they gradually become familiar with the interaction logic remains to be verified through longitudinal research. Thirdly, the sample size for individual difference analysis is relatively limited, and some variables have not shown significant relationships, which may be limited by statistical power. Fourthly, this study mainly focuses on text-based natural language interaction and has not yet addressed multimodal interactions such as speech input and visual recognition, which may significantly affect the overall gaming experience of players. Fifth, because the module order was fixed, cross-module differences may partly reflect order, practice, or fatigue effects. Therefore, scenario-specific comparisons should be interpreted cautiously. In addition, only GPT-4.1-mini was evaluated; future studies should examine whether these response-uncertainty, usability, and trust effects persist across different LLMs. Importantly, the experiment compared two bundled interaction paradigms, namely free-form LLM-mediated dialogue and preset-option scripted dialogue, rather than independently manipulating dialogue generation and input modality. Therefore, the observed cognitive-load and expression-cost differences cannot be attributed to the LLM component alone.

## VI. Conclusion

This study compares an LLM-NPC interaction paradigm combining free-form natural-language input and generated dialogue with traditional preset-option NPC interaction. The results reveal a double-edged pattern at the system level: the LLM-NPC condition showed limited improvement in overall gaming experience but was associated with higher cognitive load. On the one hand, it brings certain positive experience value to players by enhancing their perceived autonomy, and on the other hand, it offsets this advantage by increasing the expression cost, response uncertainty, and reducing system usability and trust. Further analysis shows that task scenarios significantly moderate the impact of LLM-NPC on cognitive load, particularly in modules with stronger openness such as content creation and relationship building. Meanwhile, no robust individual-difference predictors were identified after accounting for multiple comparisons. Overall, the design and application of LLM-NPC should be differentiated according to different mission scenarios, and optimized through interactive design to enhance the overall player experience, rather than simply replacing traditional NPCs.

Future research on LLM-NPC can further combine multimodal interaction forms such as speech input and visual recognition to investigate whether it can enhance the sense of presence and immersive experience while reducing the burden of expression. On the other hand, longitudinal studies that are more long-term and closely related to real commercial game scenarios can also be conducted to examine whether players gradually adapt to natural language interaction during continuous use, thereby reducing cognitive load and changing experience evaluation. In addition, in the future, efforts can be made to explore the construction of scenario based and personalized intelligent NPC design frameworks to fully unleash the potential of LLM-NPC applications in digital games.

# Appendix A. Basic Information Questionnaire

(1) Please enter your name/nickname______
(2) Please enter age______
(3) What is your gender?
A. Female B. Male C. Others
(4) Which of the following best fits your current professional background?
A. Natural Sciences B. Engineering and Technology C. Humanities and Arts D. Social Sciences E. Interdisciplinary F. High School (Science oriented) G. High School (Humanities oriented)
(5) What is your weekly gaming time?
A. Almost never played (less than 1 hour) B. Occasionally played (1-5 hours) C. Frequently played (6-10 hours) D. More played (11-20 hours) E. Heavy gamers (over 20 hours)
(6) Have you frequently played games that involve NPC interactions in the past year?
A. Yes B. No
(7) What types of games do you often play? (Multiple Choice)
A. Casual/Party Category (such as "Sugar Bean Man" and "Egg Boy Party")
B. Sandbox/Building Category (such as Minecraft and Terraria)
C. MOBA games (such as League of Legends and Honor of Kings)
D. Open world category (such as Legend of Zelda: Breath of the Wilderness, Genshin Impact)
E. Puzzle solving series (such as Portal series, Monument Valley)
F. Dramatic/visual novels (such as "Jile Disco" and "Heartbeat Literature Department")
G. Strategy/Chess (such as Civilization VI and Fire Emblem: Wind, Flower, Snow, Moon)
H. Simulation (such as "Star Dew Valley Story" and "The Sims")
I. Role playing games (such as "The Witcher 3: Wild Hunt" and "Final Fantasy VII")
J. Shooting games (such as Apex Legends and PlayerUnknown's Battlegrounds)
K. GalGame/Otome (such as "Thousand Love Flowers" and "Love and Deep Space")
(8) What is the frequency of using large language models (such as Deepseek, Qwen, ChatGPT, Gemini, etc.) in the past month?
A. Never used B. Occasionally used C. Weekly used D. Almost daily used E. Used multiple times a day
(9) What are the main scenarios in which you use the big language model? (Multiple Choice)
A. Search for information/study/ask questions B. Assist in writing/translation C. Creative generation/creation D. Chat companionship/role-playing E. Programming/work task processing
(10) To what extent do you think you are good at using large language models?
A. Very not good at B. relatively not good at C. moderately D. relatively good at E. very good at
(11) Do you have an open or positive attitude towards communicating with AI or virtual characters?
A. Completely disagree B. somewhat disagree C. generally agree D. somewhat agree E. strongly agree
(12) Do you think you have excellent language expression or writing skills?
A. Completely disagree B. somewhat disagree C. generally agree D. somewhat agree E. strongly agree
(13) Do you think you have a significant tendency towards communication anxiety or communication withdrawal?
A. Completely disagree B. somewhat disagree C. generally agree D. somewhat agree E. strongly agree
(14) Are you good at typing on a computer/phone?
A. Completely not good at B. relatively not good at C. moderately D. relatively good at E. very good at
(15) Which of the following two devices do you use more frequently?
A. Computer B. Mobile phone
(16) Please choose the trait that you are more suitable for among the following (multiple choice)
A. Neuroticism (more sensitive to emotions, more prone to tension and anxiety)
B. Extraversion (enjoys social activities and is good at socializing)
C. Openness (imaginative, exploratory, and inclusive)

D. Agreeableness (more willing to cooperate, tolerant, and sociable)
E. Conscientiousness (organized, accountable, reliable, self disciplined)

## Appendix B. Process Questionnaire

Adopting Likert 7-point scoring system

**Cognitive Load Dimension (NASA-TLX Simplified Version)**

(1) In the module just now, I put a lot of mental effort into completing the task
(2) In the module just now, I feel the pressure of time
(3) I feel like I put in a lot of effort to complete the task just now
(4) During the process of completing the task just now, I felt setbacks or difficulties

**Dimension of gaming experience**

(5) The module just now gave me a strong sense of immersion
(6) I think the interaction between the module and the character just now was very natural
(7) In the module just now, I was able to smoothly complete the interaction task with the character
(8) If possible, I am willing to continue interacting with the character I just played

**Psychological mechanism measurement**

(9) In the just interaction, I needed to spend more effort thinking about how to respond
(10) In the just interaction, I felt that I could control the direction or outcome of the interaction through my own expression
(11) I feel like I'm communicating with a character who is responding to me truthfully
(12) I clearly know what this module wants me to accomplish and how I should complete it

## Appendix C. Post Test Questionnaire

Adopting Likert 7-point scoring system

**Overall gaming experience dimension**

1. Throughout the entire trial process, I often focused on interacting with the game world and characters.
2. During the game, I pay less attention to unrelated things around me.
3. Overall, this way of interacting with NPCs makes it easier for me to enter the game state.
4. I think I am capable of handling the vast majority of tasks in this trial.
5. I think I performed well in completing these tasks.
6. This way of interacting with NPCs did not keep me in a state of "not knowing what to do" for a long time.

**Expression cost dimensions**

7. When interacting with NPCs, I need to spend a lot of effort thinking or organizing the content I want to input/express.
8. This way of interacting with NPCs makes me feel like expressing myself is a burden.
9. I often need to think extra about "how to reply to NPCs" instead of just "what to accomplish in the game".

**Autonomy dimension**

10. I feel like I can push forward the interaction with NPCs according to my own ideas.
11. I am able to express my intentions to NPCs in the way I want without being heavily restricted by the system.
12. Overall, I think this way of interacting with NPCs gives me a lot of freedom.

**Dimension of Presence**

13. I often feel like I'm interacting with a "real, responsive character".
14. The response of the characters in the game made me feel like they were "present".
15. When interacting with in-game characters, I can feel their perception and response to my expression.

**Response uncertainty dimensions**

16. When interacting with NPCs, I am not sure how to express myself in order to consistently receive the expected response.
17. In some interactions with NPCs, I am not sure if the character truly understands my intentions.

**Target clarity dimension**

18. In most modules, I am clear about what I need to accomplish.
19. In the game, I am able to understand the direction of progress for different dialogue and interactive tasks.

**System usability dimension**

20. I think this way of interacting with NPCs is generally easier to get started with.
21. I usually know quickly how to communicate with characters in the game.
22. The overall process of interacting with NPCs in this way is smooth.
23. From an operational perspective, this way of interacting with NPCs does not impose too much additional burden on me.

**Trust dimension**

24. I believe that NPCs in the game can usually understand my expression and provide useful responses.
25. I believe that the dialogue performance of NPCs in the game is reliable enough.

**Preference/Satisfaction**

26. Compared to another way of interaction, if there is a similar way of interacting with NPCs in other games, I prefer to rely on it to complete tasks.
27. I think this NPC interaction method has overall improved my gaming experience.

**Open-ended questions**

28. Which of the seven task modules do you think is the most suitable for interacting with NPCs in this way? ____
29. What do you think is the reason why this module is most suitable for interacting with NPCs among the seven task modules? ____

30. Which of the seven task modules do you think is the least suitable for interacting with NPCs in this way? ____

31. What do you think is the reason why this module is the least suitable for interacting with NPCs among the seven task modules? ____

32. Which of the seven modules would you like to experience again? ____

33. Among the seven modules, what is the reason why you are willing to experience this module again? ______

34. At which moments in the game do you feel like you are truly interacting with an NPC? ____

35. At which moments in the game are you most confused or unsure of what to do? ____

36. What do you think is the most important point for improvement in this way of interacting with NPCs? ____

37. What do you think are the most remarkable advantages and disadvantages of comparing “natural language interaction” and “preset option interaction”? ____

# Appendix D. Coding Manual

## 1. Encoding method

This study was initially coded independently by two researchers, and one researcher was responsible for coordinating disagreements and final approval. All coding personnel receive unified training before formal coding, familiarizing themselves with research questions, questionnaire structures, definitions of seven types of task modules, and coding rules in this manual. This study focuses on encoding the following types of open responses:
(1) Explanation of the reasons for the adaptability of different task modules and interaction methods;
(2) Explain why players are willing to experience a certain module again;
(3) When do players feel the most immersive and confused? Explain the reasons behind it;
(4) Subjective evaluation by players on the advantages, disadvantages, and improvement suggestions of two types of NPC interaction methods.

The coding material for this study comes from the open-ended questions of the post test questionnaire (Appendix C), mainly including text answers to questions 29, 31, 33, 34, 35, 36, and 37; The module selection results corresponding to questions 28, 30, and 32 are mainly used for grouping and contextual understanding, and are not separately used as subject coding objects.

This study uses the "smallest meaningful unit in a single answer" as the coding unit. If a participant's answer contains multiple relatively independent reasons or evaluation points, they will be divided into meaning units and encoded separately. If multiple statements revolve around the same core meaning, they are merged into one coding unit. For answers containing multiple topics, use multiple encoding. Meaningless answers (such as "nothing", "I don't know") or simply repeating the question without providing substantive content will not be included in the valid topic statistics. If the same answer involves both positive and negative directions, they should be classified under the corresponding dimensions of the topic and not cancel each other out.

## 2. Encoding process

The coding process of this study is divided into five steps:
(1) Material organization: The researcher first exports all open-ended answer texts and organizes them by question number, group (LLM-NPC group × traditional NPC group), and corresponding module. Deal with missing content that is clearly invalid (such as answering with 'none'). For texts that are semantically complete but express themselves in a colloquial and simplified manner, retain the original meaning and include it in the encoding.

(2) Pre coding and preliminary classification: Two coding personnel first pre code about 15% of the samples, independently label meaning units, and attempt to summarize the initial theme. Subsequently, the coding results were compared and discussed, with a focus on unifying the following issues: how to divide multiple parallel reasons, which statements should be considered as similar topics, which topics need to be merged or split, and how to handle vague or insufficiently informative answers. On this basis, a formal coding framework is formed and used for all subsequent samples.

(3) Formal independent coding: Two coders independently code all open-ended text according to a unified coding framework. The formal coding phase does not share results with each other. Each answer needs to complete the segmentation of meaning units and topic classification.

(4) Consistency check and divergence coordination: After the formal coding is completed, the results of two coding personnel are compared for consistency. If there is inconsistency in the topic classification of the same meaning unit, it will enter the discussion and coordination procedure. For disagreements with clear boundaries, two coders will discuss and reach a consensus based on the manual definition. For individual cases that are still difficult to determine, other members of the research team will participate in the decision-making process and make the final classification based on the principle of closest to the original semantics and least inference. The consistency test standard is set to $\alpha > 0.7$.

(5) Topic finalization and frequency statistics: After completing the coordination of differences, form the final coding table. Researchers summarize and merge synonyms for each topic, and count the frequency of topics by topic or module for presenting the results in the main text. If there are several sub items with highly similar expressions under the same category, they should be merged into one overall theme during statistics to improve the readability of the results.

**3. Example of Core Theme Definition**

The following topics are typical categories that frequently appear in the open feedback of this study, serving as important references for formal coding:

(1) Reasons for suitability/unsuitability of a certain interaction method

The reasoning process is clearer: preset options help to organize information, clarify logic, and reduce confusion in reasoning.

Reduce thinking burden: Refers to not having to organize language extensively, reducing thinking pressure, and lowering cognitive consumption.

Limited creative freedom: refers to the inability of preset options to fully express creativity, restrict imagination, or personalize expression.

Lack of challenge: Refers to the task being pushed too mechanically and directly, lacking room for expression.

Close to the relaxed atmosphere of daily life: Refers to natural language interaction that is more like real conversation and more in line with the chat scene.

Helps with emotional resonance: refers to being able to express emotions more easily, establish a sense of closeness, and feel understood.

Limited instructions are sufficient: Refers to a type of task with clear objectives, simple options are sufficient to complete, and no open input is needed.

Adding trouble and burden: Refers to the increased operational, time, or comprehension costs of natural language input.

(2) Reasons for willingness to experience again

Strong interest: Refers to the process being fun, interesting, and highly interactive.

Easy and Simple: Refers to tasks with low pressure, easy completion, and smooth experience.

Freshness/surprise: Refers to unpredictable results and more dynamic interactions.

Strong sense of participation: refers to the ability to influence the process and experience a higher level of engagement.

(3) Sense of presence and sources of confusion

Like playing/communicating with a real person: refers to feeling the other person responding, pulling, negotiating, or confronting during interaction.

Too much information, difficult to organize: refers to scattered clues, insufficient task prompts, and difficulty in integrating information.

Not sure what to do next: refers to unclear goals, insufficient feedback, and unclear direction of progress.

Unpredictable: Refers to players being unsure how to express themselves in order to achieve the expected results.
Unstable response: Refers to players experiencing hallucinations, answering irrelevant questions, or calling errors when encountering large models.

(4) Advantages, disadvantages, and improvement suggestions
Easy to operate: refers to quick and easy to pick up, click to use, and simple process.
Clear direction: refers to the system guiding clearly and the user knowing what to do.
Stable and controllable: Refers to stable feedback, predictable results, and less prone to errors.
Rigid and fixed: Refers to a single option, template based response, and lack of flexibility.
Restricting free expression: refers to the inability to express personalized ideas or strategies.
Weak sense of authenticity: Refers to a lack of genuine communication and limited immersion.
High degree of freedom: refers to the ability of users to independently organize their expression and influence the direction of interaction.
Strong immersion: Refers to a more realistic conversation and easier entry into the situation.
Strong randomness/surprise: Refers to feedback that changes frequently and is exploratory.
Time consuming and labor-intensive: Refers to the high cost of inputting, organizing language, and revising back and forth.
Lack of guidance: Refers to players not knowing how to express themselves and how to advance tasks.
Poor stability/controllability: Refers to large output fluctuations, significant understanding deviations, and unstable results.
Enhance the naturalness of response: Refers to hoping that NPC expressions are more like real people and more coherent.
Add hints and guidance: Refers to requesting the system to provide examples, keywords, task prompts, or recommended expressions.

# Appendix E. Supplementary Figures

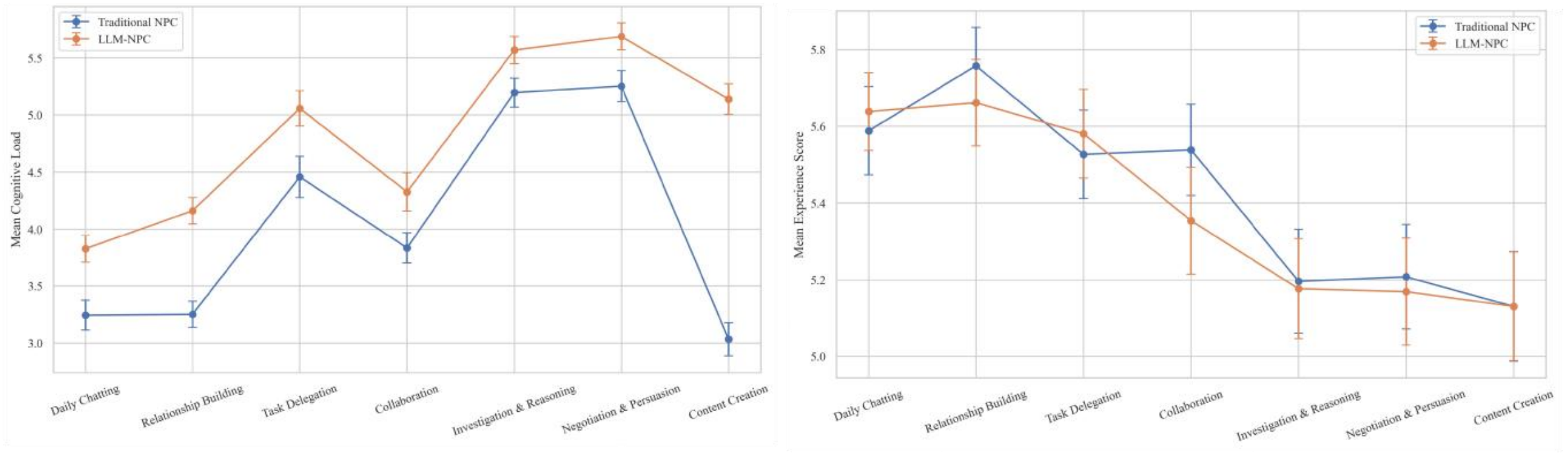

**(a)** **(b)**

**Fig. E.1.** Comparison of differences in cognitive load and gaming experience between groups
**(a)** Cognitive load **(b)** Gaming experience

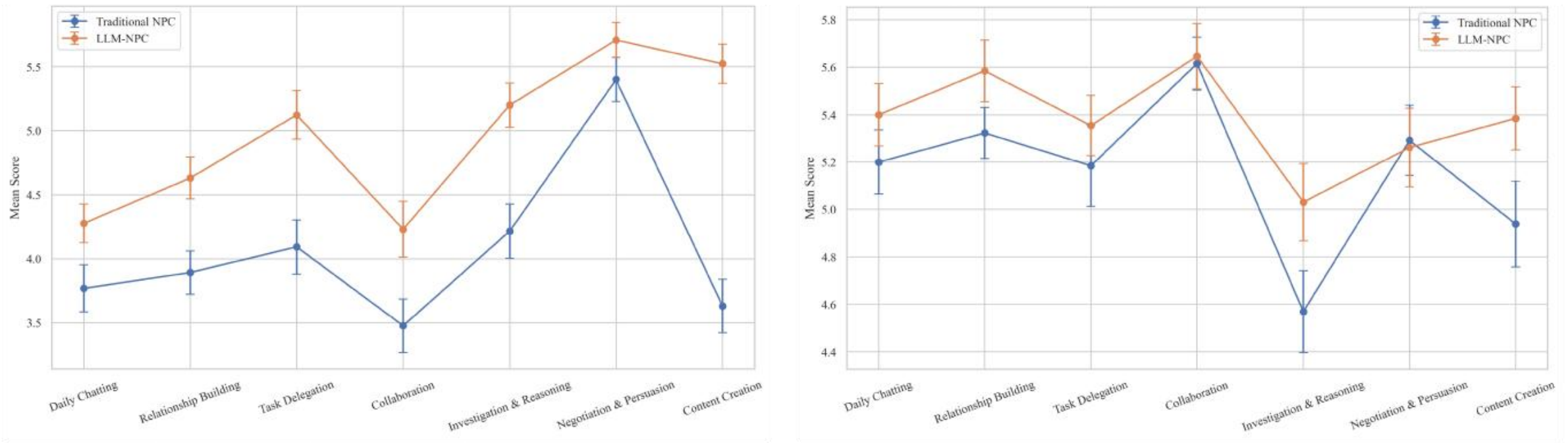

**(a)** **(b)**

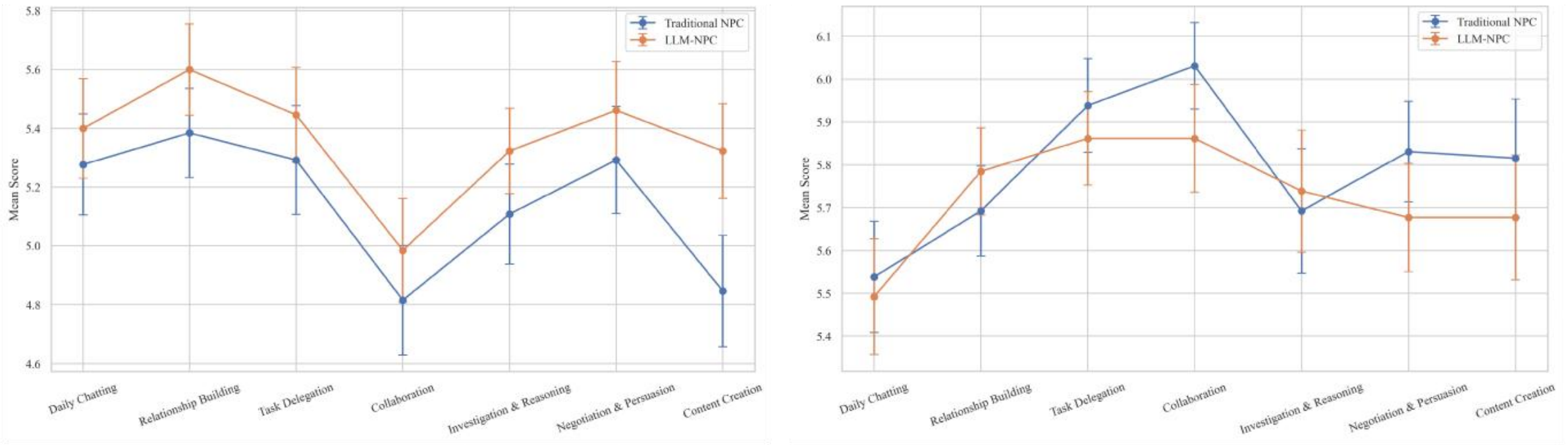

**(c)** **(d)**

**Fig. E.2.** Comparison of differences in mechanistic variables between groups
**(a)** Expression cost **(b)** Perceived antonomy **(c)** Social Presence **(d)** Target clarity

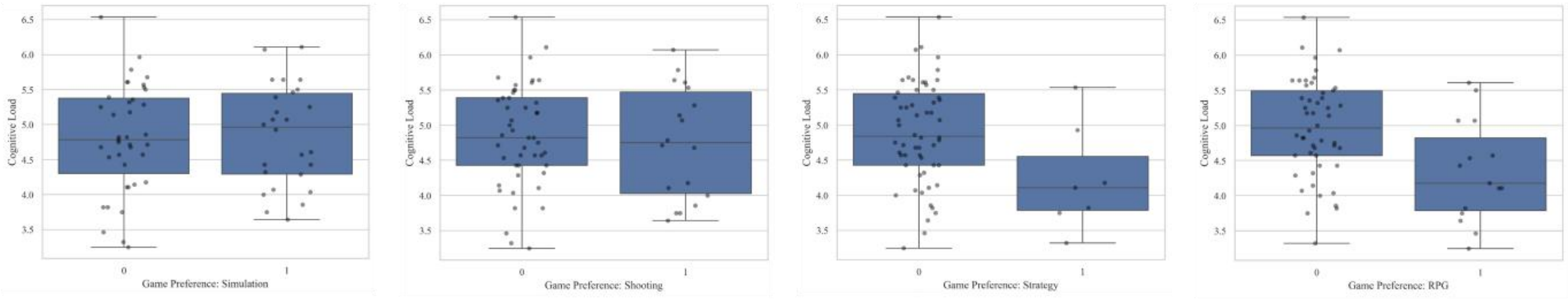

**(a)** Simulation vs. Cognitive Load **(b)** Shooting vs. Cognitive Load **(c)** Strategy vs. Cognitive Load **(d)** RPG vs. Cognitive Load

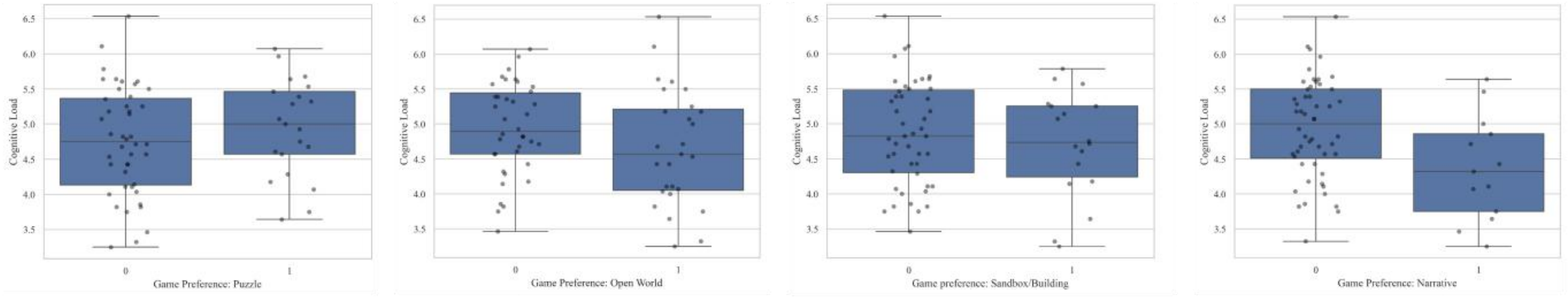

**(e)** Puzzle vs. Cognitive Load **(f)** Open World vs. Cognitive Load **(g)** Sandbox/Building vs. Cognitive Load **(h)** Narrative vs. Cognitive Load

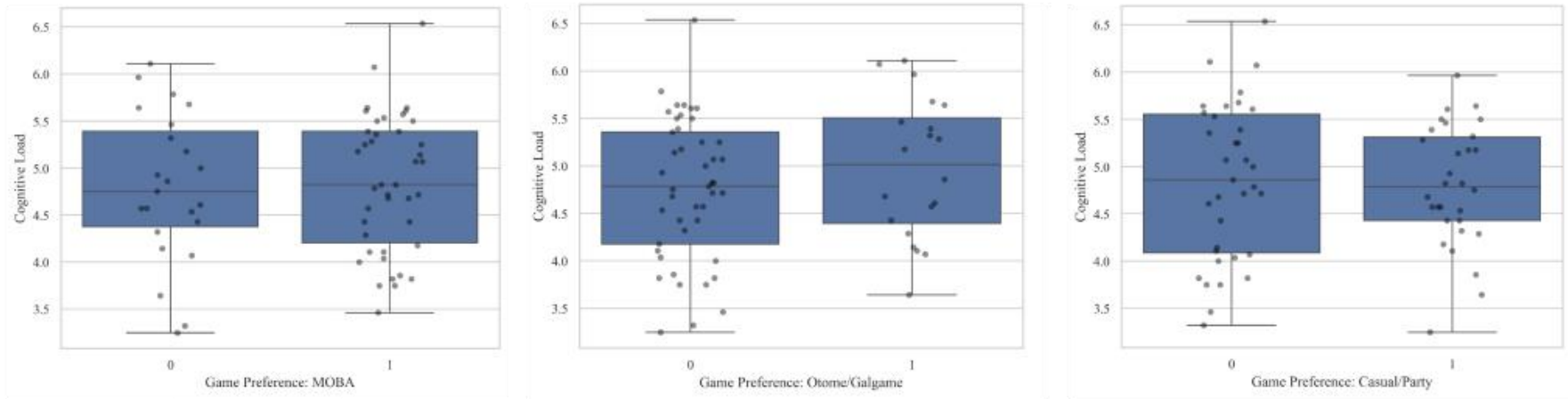

**(i)** MOBA vs. Cognitive Load **(j)** Otome/Galgame vs. Cognitive Load **(k)** Casual/Party vs. Cognitive Load

**Fig. E.3.** Differences in cognitive load of player game preferences in LLM-NPC interactions

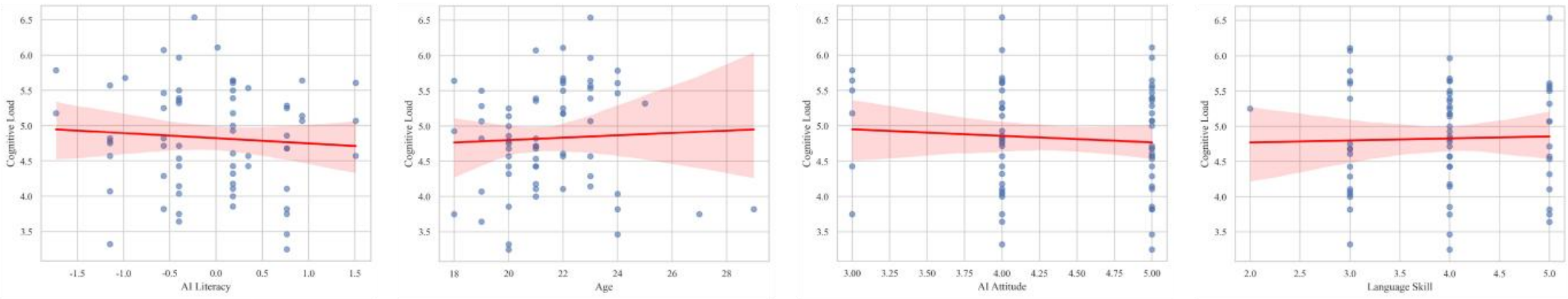

**(a)** AI Literacy vs. Cognitive Load **(b)** Age vs. Cognitive Load **(c)** AI Attitude vs. Cognitive Load **(d)** Language Skill vs. Cognitive Load

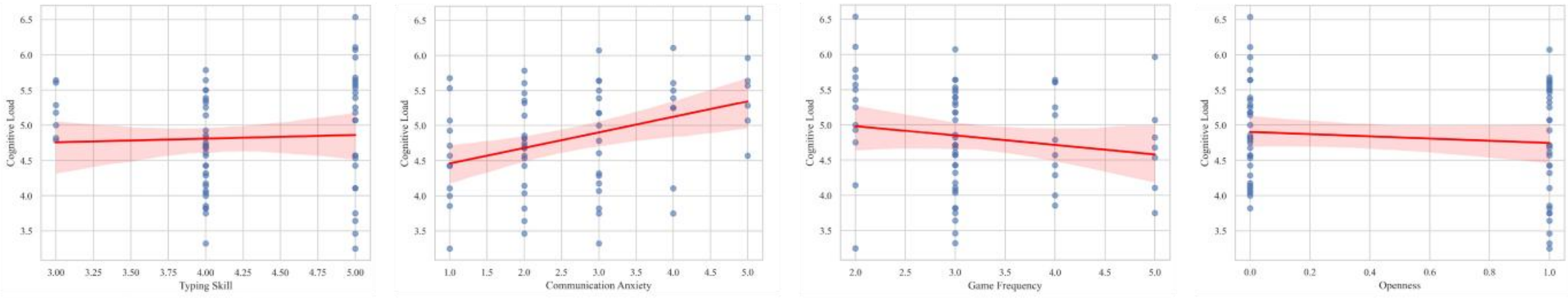

**(e)** Typing Skill vs. Cognitive Load **(f)** Communication Anxiety vs. Cognitive Load **(g)** Game Frequency vs. Cognitive Load **(h)** Openness vs. Cognitive Load

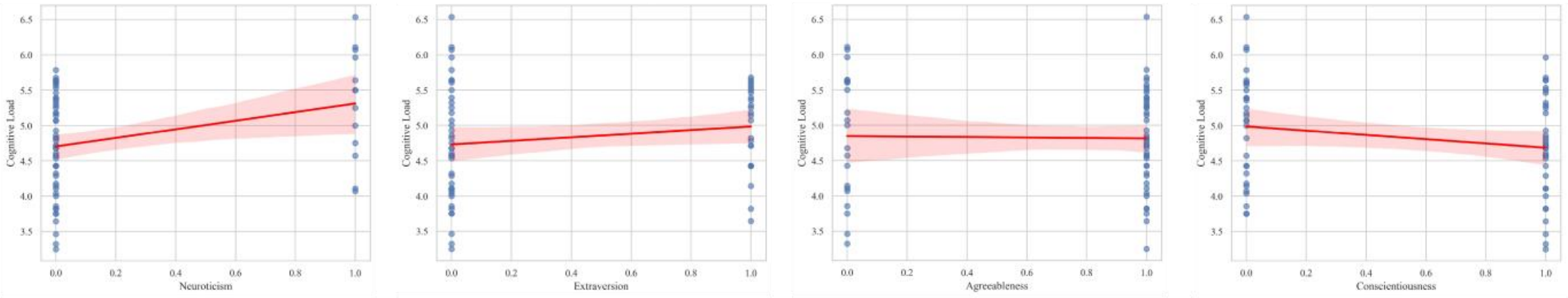

**(i)** Neuroticism vs. Cognitive Load **(j)** Extraversion vs. Cognitive Load **(k)** Agreeableness vs. Cognitive Load **(l)** Conscientiousness vs. Cognitive Load

**Fig. E.4.** Analysis results of multiple linear regression model